\documentclass[10pt]{article}
\usepackage{setspace}
\usepackage{extsizes}

\usepackage[left = 2cm, right=2cm, top=3cm, bottom=3cm]{geometry}
\usepackage[pdftex]{graphicx}
\usepackage{cite}
\usepackage{bm}
\usepackage{float}
\usepackage{amsmath,amssymb,latexsym}
\usepackage{color}
\usepackage[T1]{fontenc}
\usepackage{wasysym}
\usepackage{siunitx}
\usepackage{relsize}
\usepackage{tikz}
\usepackage{lineno, blindtext}
\usepackage{authblk}
\usepackage[symbol]{footmisc}
\usepackage{lineno}
\usepackage{multirow}
\usepackage{chemfig}
\usepackage[version=4]{mhchem}
\usepackage{pdfpages}

\begin{document}
\renewcommand{\figurename}{Fig.}
\title{\color{blue}\textbf{Role of microgel stiffness in particle self-assembly and suspension rheology across the lower consolute solution temperature}}
\author[1, $\dagger$]{Chandeshwar Misra}
\affil[1]{\textit{Soft Condensed Matter Group, Raman Research Institute, C. V. Raman Avenue, Sadashivanagar, Bangalore 560 080, INDIA}}
\author[1, $\ddagger$]{Sanjay Kumar Behera}
\author[1, $\ddagger$]{Sonali Vasant Kawale}
\author[1,*]{Ranjini Bandyopadhyay}
\date{\today}

\footnotetext[2]{misra@pc.rwth-aachen.de}
\footnotetext[1]{Corresponding Author: Ranjini Bandyopadhyay; Email: ranjini@rri.res.in}
\maketitle
\begin{abstract}We synthesize thermoresponsive poly($N$-isopropylacrylamide) (PNIPAM) colloidal microgel particles of different stiffnesses by controlling the concentration of crosslinker in a one-pot synthesis method. We employ oscillatory rheology and cryogenic scanning electron microscopy to study the temperature and stiffness-induced mechanical properties and microscopic structures of dense aqueous suspensions of the synthesized PNIPAM microgels. Using Fourier transform infrared (FTIR) spectroscopy, we show that particle hydrophobicity increases with increasing suspension temperature and decreasing particle stiffness. Our zeta potential measurements of soft PNIPAM particles and those of intermediate stiffnesses demonstrate that these particles are electrostatically unstable and prone to aggregation even at temperatures below the lower consolute solution temperature (LCST). In contrast, stiff PNIPAM particles in dilute aqueous suspensions are electrostatically stabilized at all temperatures explored in this study. Interestingly, our frequency and strain amplitude sweep rheology experiments reveal that the linear viscoelastic moduli and yield stresses of all the PNIPAM suspensions increase when the temperature is raised above the LCST. Combining cryogenic scanning electron microscopy (cryo-SEM) and rheology, we demonstrate that dense suspensions of soft PNIPAM microgels show a gel-liquid-gel transition with increase in temperature across the LCST. Suspensions of stiff particles, in contrast, exhibit a glass-glass transition under the same temperature sweep conditions and do not pass through an intermediate liquid state. Our observations reveal that the temperature-induced phase transformations in dense PNIPAM suspensions depend sensitively on the stiffness of the constituent particles and can be explained by considering amphiphilicity-driven morphological changes in the suspension microstructures over the experimental temperature range. We can therefore achieve a variety of dense suspension phases with unique properties by tuning just the particle stiffness and suspension temperature. The results reported here can be generalized to dense suspensions of a range of thermoresponsive microgel particles and have broad implications in the development of multifunctional systems such as soft robots and micro-grippers.
\end{abstract}

\renewcommand{\thesection}{\Roman{section}} 
\renewcommand{\thesubsection}{\thesection.\Roman{subsection}}
    \definecolor{aqua}{rgb}{0.0, 1.0, 1.0}
	\definecolor{brass}{rgb}{0.71, 0.65, 0.26}
	\definecolor{junglegreen}{rgb}{0.16, 0.67, 0.53}
	\definecolor{indigo(web)}{rgb}{0.29, 0.0, 0.51}
	\definecolor{alizarin}{rgb}{0.82, 0.1, 0.26}
	\definecolor{red(ryb)}{rgb}{1.0, 0.15, 0.07}
	\definecolor{black}{rgb}{0.0, 0.0, 0.0}
	\definecolor{darkcandyapplered}{rgb}{0.64, 0.0, 0.0}
	\definecolor{darkred}{rgb}{0.55, 0.0, 0.0}
	\definecolor{darkolivegreen}{rgb}{0.33, 0.42, 0.18}
	\definecolor{internationalkleinblue}{rgb}{0.0, 0.18, 0.65}
	\definecolor{olivedrab(web)(olivedrab3)}{rgb}{0.42, 0.56, 0.14}
	\definecolor{saddlebrown}{rgb}{0.55, 0.27, 0.07}
	\definecolor{purple(munsell)}{rgb}{0.62, 0.0, 0.77}
	\definecolor{ufogreen}{rgb}{0.24, 0.82, 0.44}
	 \definecolor{purpleheart}{rgb}{0.41, 0.21, 0.61}
		\definecolor{blue(ryb)}{rgb}{0.01, 0.2, 1.0}
	\newcommand{\bt}{\textcolor{brass}{$\blacktriangle$}}
	\newcommand{\btbullet}{\textcolor{brass}{\large$\bullet$}}
	\newcommand{\brass}{\textcolor{brass}{\small$\ocircle$}}
	\newcommand{\BL}{\textcolor{black}{\small$\blacksquare$}}
	\newcommand{\blbullet}{\textcolor{black}{\large$\bullet$}}
	\newcommand{\darkred}{\textcolor{darkred}{$\blacksquare$}}
	\newcommand{\wine}{\textcolor{darkred}{\large$\bullet$}}
	\newcommand{\DR}{\raisebox{2pt}{\tikz{\draw[-,black!40!darkcandyapplered,solid,line width = 2.5pt](0,0) -- (5mm,0);}}}
	\newcommand{\hollow}{\textcolor{black}{$\square$}}
	\newcommand{\JG}{\textcolor{junglegreen}{\large$\varhexagon$}}
	\newcommand{\cyan}{\textcolor{junglegreen}{$\bigstar$}}
	 \newcommand{\cyanbullet}{\textcolor{junglegreen}{\large$\bullet$}}
	\newcommand{\green}{\textcolor{darkolivegreen}{$\bigstar$}}
	\newcommand{\bluepentagon}{\textcolor{internationalkleinblue}{\large$\pentagon$}}
	\newcommand{\purplepentagon}{\textcolor{purple(munsell)}{\small$\triangle$}}
	\newcommand{\purbulletsolid}{\textcolor{purpleheart}{\large$\bullet$}}
	\newcommand{\brown}{\textcolor{saddlebrown}{\large$\diamond$}}
	\newcommand{\olivedot}{\textcolor{olivedrab(web)(olivedrab3)}{$\square$}}
	\newcommand{\prline}{\raisebox{2pt}{\tikz{\draw[-,black!40!purple(munsell),solid,line width = 2.5pt](0,0) -- (5mm,0);}}}
	\newcommand{\wineline}{\raisebox{2pt}{\tikz{\draw[-,black!40!darkred,solid,line width = 2.5pt](0,0) -- (5mm,0);}}}
	\newcommand{\gline}{\raisebox{2pt}{\tikz{\draw[-,black!40!junglegreen,solid,line width = 2.5pt](0,0) -- (5mm,0);}}}
	\newcommand{\bline}{\raisebox{2pt}{\tikz{\draw[-,black!40!blue(ryb),solid,line width = 2.5pt](0,0) -- (5mm,0);}}}
	\newcommand{\indigo}{\textcolor{indigo(web)}{\large$\bullet$}}
	\newcommand{\blueline}{\raisebox{2pt}{\tikz{\draw[-,black!40!internationalkleinblue,solid,line width = 2.5pt](0,0) -- (5mm,0);}}}
	\newcommand{\oliveline}{\raisebox{2pt}{\tikz{\draw[-,black!40!ufogreen,solid,line width = 2.5pt](0,0) -- (5mm,0);}}}
	\newcommand{\candyred}{\textcolor{darkcandyapplered}{\large$\bullet$}}
	\newcommand{\blhex}{\textcolor{black}{\large$\varhexagon$}}
	\newcommand{\brsquare}{\textcolor{brass}{\small$\square$}}
	\newcommand{\brcircle}{\textcolor{brass}{\large$\circ$}}
	\newcommand{\prcircle}{\textcolor{purple(munsell)}{\large$\circ$}}
	\newcommand{\olcircle}{\textcolor{junglegreen}{\large$\circ$}}
	\newcommand{\blcircle}{\textcolor{blue(ryb)}{\large$\circ$}}
	\newcommand{\rdcircle}{\textcolor{darkcandyapplered}{\large$\circ$}}
	\newcommand{\brtria}{\textcolor{brass}{\small$\triangle$}}
	\newcommand{\brdown}{\textcolor{brass}{\small$\triangledown$}}
	\newcommand{\brdiamond}{\textcolor{brass}{\large$\diamond$}}
	\newcommand{\bltriangle}{\textcolor{internationalkleinblue}{$\triangleleft$}}
	\newcommand{\blrhd}{\textcolor{internationalkleinblue}{\large$\triangleright$}}
	\newcommand{\blcirc}{\textcolor{internationalkleinblue}{\large$\circ$}}
	\newcommand{\blpentagon}{\textcolor{internationalkleinblue}{$\pentagon$}}
	\newcommand{\bltri}{\textcolor{internationalkleinblue}{\small$\triangle$}}
	\newcommand{\blsq}{\textcolor{internationalkleinblue}{\small$\square$}}
	\newcommand{\rddiam}{\textcolor{darkcandyapplered}{\large$\diamond$}}
	\newcommand{\rdhexagon}{\textcolor{darkcandyapplered}{\small$\varhexagon$}}
     \newcommand{\wnbullet}{\textcolor{darkred}{\large$\bullet$}}
	 \newcommand{\blsquare}{\textcolor{black}{\large$\bullet$}}
\vspace{0.1mm}
\renewcommand \thesection{\arabic{section}}
\renewcommand \thesubsection{\arabic{section}.\arabic{subsection}}
\section{Introduction}
Colloidal suspensions have been studied extensively and systematically because of their importance in modeling phenomena such as the glass transition \cite{C_A_Angell_2000,J_Mattsson_2009,SanjayPRM,Silvia_Franco_2021,Van_der_Scheer_2017}, and for their applicability in therapeutics and in the manufacture of food, paints, sensors etc \cite{Romeo2010,Senff_H_1999,SanjayPRM,Lyon_L_2012,DONG1991141,Nolan2008-hp,Islam_2014}. Soft deformable colloidal particles, which can be produced by crosslinking polymer particles immersed in a solvent, offer an advantage over many hard colloidal particles due to their significant swelling and deswelling properties in response to changes in external stimuli such as temperature, pH and solvent composition \cite{Romeo2010,S_Hirotsu,M_H_Dufresne_2004,Lopez-Leon_2007}.  
\paragraph{}
In recent years, researchers have comprehensively investigated the complex phase behavior, inter-particle interactions and jamming dynamics of dense suspensions of soft thermoresponsive poly({\it N}-isopropylacrylamide) (PNIPAM) colloidal microgel particles \cite{Romeo2010,Yodh_RPP_2014, Fernandez-Nieves_softmatter_2012, Scotti_PNAS_2016, Zhang_Nature_2009, Cloitre_softmatter_2009}. In aqueous suspension, these particles undergo a reversible volume phase transition above a lower consolute solution temperature (LCST) of $\approx$ $34^\circ$C \cite{Romeo2010,Senff_H_1999}. Below the LCST, PNIPAM particles swell due to the absorption of water from the surrounding medium. Above the LCST, however, they become hydrophobic and expel water, resulting in a dramatic shrinkage in their sizes. The properties of microgel suspensions can therefore be tuned simply by controlling the temperature \cite{Lyon_L_2012,Senff_H_1999,Saunders_ACIS_2008,Han_PRE_2008,Alsayed_Science_2005}. Their thermoreversible nature makes PNIPAM suspensions ideal candidates for drug delivery and in the preparation of strong gels while designing soft bio-mimetic devices \cite{DONG1991141, Calvert_advmater_2009}. Nanocomposite PNIPAM microgels have been shown to exhibit highly robust mechanical properties. For instance, it has been demonstrated recently that poly(acrylamide) nanocomposite microgels can resist compressive and tensile stresses of the order of megapascals \cite{Hu_Materials-design_2019}. The excellent mechanical resilience and thermoresponsivity of microgel suspensions make them suitable for use as multifunctional materials in biomedical engineering \cite{Calvert_advmater_2009, Nicholas_advmatter_2006}.
\paragraph{}
Recent research has established the length-scale dependent rheology of PNIPAM suspensions \cite{C_MISRA_JCIS_2022}. Using a rheo-dielectric technique, the authors of this study reported that while dynamics of PNIPAM chains slow down under large amplitude oscillatory shears due to the entanglement of polymer chains at the nanoscale, the bulk stress relaxation speeds up simultaneously due to the fragmentation of self-assembled PNIPAM clusters under large deformations. It has been demonstrated experimentally that PNIPAM particle stiffness can be altered by varying the concentration of the crosslinker ($N,N^{\prime}$-methylenebisacrylamide (MBA)) while keeping the concentrations of the monomer  ({\it N}-isopropylacrylamide) and the surfactant (sodium dodecyl sulfate -SDS) fixed during the synthesis process \cite{McPhee_JCIS_1993,Senff_CPS_2000,CMisra2020}. A recent study reported that the stiffness of PNIPAM particles can be controlled by changing poly(acrylic acid) (PAAc) concentrations \cite{Franco_polym_2022}. The dependence of suspension rheology on particle stiffness was recently tested experimentally for dense suspensions of PNIPAM particles that were crosslinked with glutaraldehyde \cite{D_Maier_2022}. The authors of this study used a schematic mode coupling theory (MCT) to model the linear rheology of the samples below the LCST. According to another study, a suspension of relatively stiff ionic microgel particles below the LCST shows a transition from liquid to crystal to glass with increase in volume fraction \cite{Fernandez-Nieves_softmatter_2012}. In contrast, suspensions of particles of intermediate stiffnesses transform from liquid to glass without going through an intermediate crystal phase. Interestingly, suspensions of the softest particles always remained liquid-like across the range of volume fractions explored in the study. Iyer {\it et al.} \cite{Iyer_ACIE_2009} studied the self-healing properties of microgel colloidal crystals in the presence of particle-size irregularity and reported that self-assemblies of these particles are defect-tolerant due to the softness of the interparticle interactions and the ability of the particles to dissipate defect energies over long length scales. They reported that microgels exhibit characteristics of both polymeric and colloidal systems. It has been reported that the rheological properties of PNIPAM suspensions depend on the synthesis protocol, average particle size, size polydispersity and particle crosslinking density \cite{SanjayPRM,Senff_CPS_2000,CMisra2020}. A higher particle crosslinking density leads to a higher interparticle force, thereby increasing the plateau value of the elastic modulus of the suspension \cite{Senff_CPS_2000}. In contrast, the easy penetration of the long dangling polymer chains present on the surfaces of particles with lower crosslinking densities results in higher yield stresses.   
\paragraph{}
The temperature-dependent phase behavior of dense aqueous PNIPAM suspensions has been widely studied in the literature \cite{J_Mattsson_2009,Senff_H_1999,Romeo2010}. Using dynamic light scattering and rheological measurements, J. Mattsson {\it et al.} \cite{J_Mattsson_2009} studied the glass transition dynamics of dense suspensions of microgels constituted by interpenetrated networks of PNIPAM and PAAc (polyacrylic acid). They reported that these suspensions mimic molecular glass formers, with suspension fragility decreasing with increasing microgel stiffness. Romeo {\it et al.} have reported that the glassy state formed by temperature and pH-responsive PNIPAM suspensions at low temperature liquefies at the LCST before solidifying into a volume-spanning colloidal gel-like state at higher temperatures \cite{Romeo2010}. The authors explained their results by considering that the interparticle interaction between PNIPAM particles changes from repulsive to attractive across the LCST. The intermediate liquid state was understood to result from a reduction in effective volume fraction due to sudden collapse in the sizes of PNIPAM particles at the LCST. Stiffness and temperature-induced phase changes in PNIPAM suspensions have been widely investigated  in separate studies \cite{J_Mattsson_2009,Franco_polym_2022,Romeo2010,Senff_H_1999,Fernandez-Nieves_softmatter_2012,Senff_CPS_2000}. To the best of our knowledge, an extensive study on the effects of simultaneous control of both particle stiffness and suspension temperature on the phase behavior and mechanical properties of dense aqueous PNIPAM suspensions has never been reported in the literature. In the present study, we systematically demonstrate that dense aqueous PNIPAM suspensions can exhibit a rich variety of phases due to controlled changes in suspension temperature and particle stiffness while keeping the effective volume fraction fixed at a temperature below the LCST.
\paragraph{}
We synthesize PNIPAM particles of different stiffnesses by varying the concentration of the crosslinker in a free radical precipitation polymerization synthesis method. The particles are then suspended in an aqueous medium to form dense suspensions. Mechanical properties of these suspensions, such as their linear viscoelastic moduli and yield stresses, are studied in oscillatory rheology experiments. Particle stiffness is quantified in terms of a maximum swelling ratio. We report that for a fixed volume fraction below the LCST, the elasticity of a dense aqueous PNIPAM suspension increases with increase in particle stiffness. The viscoelastic moduli of the suspensions exhibit non-monotonic change with increasing temperature. Both moduli decrease significantly as the LCST $\approx 33{^\circ}$C is approached and then increase steadily as the temperature is raised above the LCST. This counter-intuitive increase is driven by hydrophobicity-induced clustering and is in agreement with the results of the study by Romeo {\it et al.} \cite{Romeo2010}. We estimate the hydrophobicities and zeta potentials of PNIPAM particles of different stiffnesses using Fourier transform infrared (FTIR) spectroscopy and an electroacoustic  technique, respectively. These measurements allow us to study the competition between electrostatic and hydrophobic inter-particle interactions between PNIPAM particles constituting the dense aqueous suspensions studied here. Finally, we correlate the macroscopic mechanical behavior (bulk rheology) of these dense suspensions with their microscopic structures by directly visualizing the cryo-frozen samples under a cryogenic scanning electron microscope (cryo-SEM). Cryo-SEM images of dense suspensions of soft PNIPAM particles, prepared at temperatures below and above the LCST, display three-dimensional network structures. For suspensions prepared at the LCST, we note the coexistence of liquid-like and highly porous gel-like phases. Suspensions of stiff particles, in contrast, always assemble in a disordered glassy state over the entire temperature range. {Our work demonstrates that the balance between attractive hydrophobic interactions and electrostatic repulsion plays an extremely important role in determining the phase behavior of dense suspensions of PNIPAM particles. Furthermore, our results clearly indicate that the temperature-dependent phase behavior of dense PNIPAM suspensions can be tailored effectively by controlling the degree of particle crosslinking and therefore the stiffnesses of individual PNIPAM particles. We believe that a clearer understanding of the microstructural details and bulk rheology of dense suspensions constituted by particles of controlled stiffnesses will trigger systematic research on their use as multifunctional materials \cite{Calvert_advmater_2009}.

\section{Material and Methods}   
\subsection{Synthesis of PNIPAM particles} 	
Poly($N$-isopropylacrylamide) (PNIPAM) particles of different stiffnesses were synthesized by varying the crosslinker concentration in a free radical precipitation polymerization process. All the chemicals were purchased from Sigma-Aldrich and used as received without any further purification. In the polymerization reaction, 7.0 g $N$-isopropylacrylamide (NIPAM) ({\it 99$\%$}), predetermined weights of the crosslinker $N,N'$-methylenebisacrylamide (MBA) ({\it 99.5$\%$}) and 0.03 g SDS were dissolved in 470 ml Milli-Q water (Millipore Corp.) in a three-necked round-bottomed (RB) flask attached with a reflux condenser, a magnetic stirrer with heating (Heidolph), a platinum sensor and a nitrogen gas (N$_2$) inlet/outlet. The solution was then stirred at 600 RPM and purged with N$_2$ gas for 30  min to remove oxygen present in the solution. Dissolution of 0.28 g potassium persulphate (KPS) (99.9$\%$) in 30 ml Milli-Q water at 70$^\circ$C initiated a polymerization reaction that was allowed to proceed for 4 h while stirring at 600 RPM. The suspension was then cooled down to room temperature. PNIPAM particles of different stiffnesses were obtained by varying the concentration of the crosslinker $N,N'$-methylenebisacrylamide (MBA) between 1\% and 25\% of the NIPAM monomer mass. The suspension was then purified by four successive centrifugations and re-dispersions at rotational speeds of 15,000-50,000 RPM for 60 min to remove SDS, the remaining monomers, oligomers, and impurities. We used higher centrifugation speeds for particles synthesized at lower MBA concentrations and vice-versa. After the centrifugation, the supernatant was removed and the remaining sample was dried by evaporating the water. A fine powder was prepared by grinding the dried particles using a mortar and pestle. The synthesis conditions and the swelling properties of the synthesized particles are tabulated in Table 1.
\begin{table*}
			\begin{center}
				\begin{tabular}{ |c|c|c|c|c|c|c|c| }
					\hline

			$NIPAM$ & $MBA$ & $KPS$ & $SDS$ & $\alpha$ & $c_{p}$ & $\phi_{eff}$ & $c$ \\
				(g) & $($\%$~of~NIPAM)$ & $(g)$ & $(g)$ & $(d_{fullswell}/d_{fullshrunk})$ & $(wt.\%)$ &  & $(wt.\%)$ \\
				\hline
				7 & $1.0$ & $0.28$ & $0.03$ & ${\approx} 3.36$ & $3.63$ & $1.5$ & $5.44$ \\
				\hline
				7 & $2.5$ & $0.28$ & $0.03$ & ${\approx} 2.56$ & $5.06$ & $1.5$ & $7.59$ \\
				\hline
				7 & $5.0$ & $0.28$ & $0.03$ & ${\approx} 2.00$ & $7.57$ & $1.5$ & $11.35$ \\
				\hline
				7 & $10.0$ & $0.28$ & $0.03$ & ${\approx} 1.66$ & $10.87$ & $1.5$ & $16.30$ \\
				\hline
				7 & $15.0$ & $0.28$ & $0.03$ & ${\approx} 1.51$ & $17.00$ & $1.5$ & $25.5$ \\
				\hline
				7 & $20.0$ & $0.28$ & $0.03$ & ${\approx} 1.41$ & $21.63$ & $1.5$ & $32.44$ \\
				\hline
				7 & $25.0$ & $0.28$ & $0.03$ & ${\approx} 1.33$ & $28.82$ & $1.5$ & $43.23$ \\
				\hline
				\end{tabular}
				\caption{Synthesis conditions for the preparation of PNIPAM particles by the one-pot polymerization method and characterization of the swelling properties (stiffnesses) of the synthesized particles. All PNIPAM particles were prepared in $500$ ml volume. $\alpha$, $c_{p}$, $\phi_{eff}$, $c$ are, respectively, the maximum swelling ratio of the PNIPAM particle obtained from dynamic light scattering experiments (section 1 of the supplementary information), polymer concentration inside each PNIPAM particle, effective volume fraction and PNIPAM particle concentration in aqueous suspension obtained from rheological measurements (section 3 of the supplementary information) at 25$^\circ$C.}
				\label{table:Synthesis}
			\end{center}
		\end{table*}
\subsection{Dynamic light scattering}
Dynamic light scattering (DLS) experiments were performed to quantify the average sizes and thermoresponsivity of PNIPAM particles of different stiffnesses in dilute suspensions. The DLS experiments were carried out using a Brookhaven Instruments Corporation (BIC) BI- 200SM spectrometer. The details of the setup are described elsewhere \cite{Debasish_Soft_Matter_2014}. A water circulating temperature controller unit (Polyscience Digital) was used to maintain temperature between 16$^{\circ}$C and 55$^{\circ}$C. A Brookhaven Instruments Corporation digital autocorrelator, BI-9000AT, was used to compute the intensity autocorrelation function, $g^{(2)} (q, t)$, of the scattered light. The intensity autocorrelation function is defined as $g^{(2)}(q,t) = \frac{<I(q,0)I(q,t)>}{<I(q,0)>^{2}} =  1+ A|g^{(1)}(q,t)|^{2}$, where $I(q,t)$, $g^{(1)}(q,t)$ and $A$ are the scattered intensity at a particular scattering wave vector $q$ and a delay time $t$, the normalized electric field autocorrelation function and the coherence factor, respectively. The angular bracket $<>$ in the expression represents an average over time. The scattering wave vector $q=(4\pi n/\lambda)\sin(\theta/2)$, where $\theta$, $n$ and $\lambda$ are, respectively, the scattering angle (90$^{\circ}$ used here), the refractive index of the medium ($\approx$ 1.33) and the laser wavelength (532 nm in our case). 
\paragraph{}
For a suspension of size-polydisperse particles, the decays of the normalized autocorrelation function $C(t)= \frac{g^{(2)}(q,t)-1}{A}$ are fitted to stretched exponential functions of the form $C(t)=\big[e^{-(t/\tau)^{\beta}}\big]^{2}$. The stretching exponent $\beta$, a fitting parameter, is used to calculate the mean relaxation time of the particles in suspension using the relation $<$$\tau$$>$$=$$(\frac{\tau}{\beta})\Gamma(1/\beta)$, where $\Gamma$ is the Euler Gamma function \cite{SanjayPRM,CMisra2020}. The mean hydrodynamic diameter $<$$d_{H}$$>$ of the particles is estimated using the Stokes-Einstein relation  $<$$d_{H}$$>$$=\frac{k_{B}T<\tau>q^{2}}{3\pi\eta}$, where $k_{B}$, $T$ and $\eta$ are the Boltzmann constant, the absolute temperature and viscosity of the solvent, respectively \cite{SanjayPRM,B_J_Berne_1976,A_Einstein_1905}. Fig. S1 of the supplementary information shows the temperature-dependent average hydrodynamic diameters, $<d_{H}>$, of PNIPAM particles synthesized using different crosslinker (MBA) concentrations in aqueous suspensions.
\begin{figure}[!t]
\centering
\includegraphics[width=3.5in]{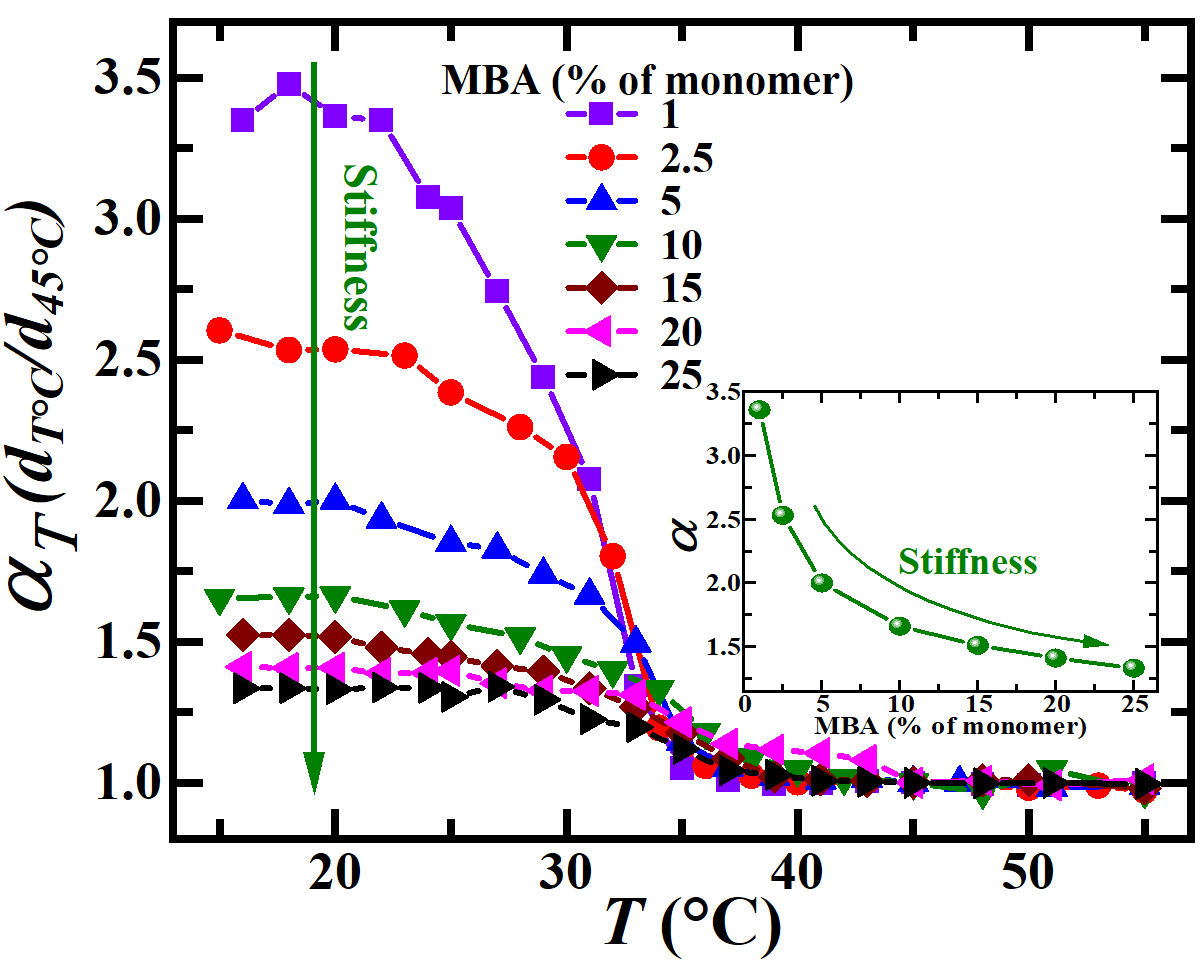}
\caption{Temperature-dependent swelling ratios, $\alpha_{T}=d_{T^{\circ}C}/d_{45^{\circ}C}$, where $d_{T^{\circ}C}$ and $d_{45^{\circ}C}$ are the average hydrodynamic diameters of PNIPAM particles at temperatures $T$ and $45^{\circ}$C respectively, synthesized with different crosslinker (MBA) concentrations in dilute aqueous suspensions. The inset shows the dependence of maximum swelling ratios, $\alpha$, of these particles on the concentration of the crosslinker used during particle synthesis. The arrows point towards increasing stiffness of PNIPAM particles.}	
\label{DLS}
\end{figure}
\paragraph{}
Very dilute suspensions ($\phi_{eff} \approx 10^{-4}$) were prepared by adding dried PNIPAM powder in Milli-Q water. A glass cuvette, filled with 5.0 ml of the suspension, was placed in the sample holder of the DLS setup. We measured the hydrodynamic diameters $<$$d_{H}$$>$ of the thermoresponsive PNIPAM particles within a temperature range between $16^{\circ}$C and $55^{\circ}$C at intervals of $2^{\circ}$C. We quantified the swelling behavior of PNIPAM particles by measuring their maximum swelling ratios, $\alpha$ = $d_{20^{\circ}C}/d_{45^{\circ}C}$, where $d_{20^{\circ}C}$ and $d_{45^{\circ}C}$ refer to the average hydrodynamic diameters of the particles at $20$$^{\circ}$C and $45$$^{\circ}$C where they are fully swollen and fully shrunken/ collapsed, respectively.
\paragraph{}
Fig. \ref{DLS} shows the temperature-dependent swelling ratios, $\alpha_{T}=d_{T^{\circ}C}/d_{45^{\circ}C}$, where $d_{T}$ is the average hydrodynamic diameter of the particles at temperature $T$. Due to the higher osmotic pressure inside soft particles, the polymer networks in PNIPAM particles synthesized using low crosslinker concentration undergo rapid expansion by incorporating water \cite{J_Mattsson_2009,Senff_CPS_2000}. It has been reported that the equilibrium size of PNIPAM particles is determined by the balance between temperature-dependent osmotic pressure and elastic stresses in the networks \cite{Fernandez-Nieves_softmatter_2012}. We see from Fig. \ref{DLS} that the volume phase transition of these particles at the LCST becomes more continuous with increase in crosslinker concentration. This aspect is illustrated in Fig. S2 where we have plotted the rate of change of $\alpha_{T}$ with {\it T}, $d\alpha_{T}/dT$, within a narrow temperature range near the LCST. In this plot, $d\alpha_{T}/dT$ is obtained by linearly fitting the portion of the data in Fig. \ref{DLS} where the particle diameters show an approximately linear collapse. The decrease in $d\alpha_{T}/dT$ with increasing crosslinker concentration i.e. increasing PNIPAM particle stiffness, confirms that the volume phase transition near the LCST becomes more continuous with increasing stiffness of individual particles.\\

Our DLS data reveals that the PNIPAM particles synthesized by us have maximum swelling ratios, $\alpha$, that range between 3.36 (soft particles) to 1.33 (stiff particles) as we varied the crosslinker (MBA) concentrations between 1.0 to 25.0\% of the NIPAM monomer mass. The values of $\alpha$ corresponding to the various concentrations of the crosslinker (MBA) are listed in Table 1 and are plotted in the inset of Fig. \ref{DLS}. We see from the inset of Fig. \ref{DLS} that the maximum swelling ratios, $\alpha$, of PNIPAM particles in aqueous suspension decrease with increase in crosslinker (MBA) concentration, signaling an increase in particle stiffness. In this study, we have designated PNIPAM particles with maximum swelling ratios $\alpha$ $\ge$ 2.56 as soft particles, particles with  1.51 $\le$ $\alpha$ $\le$ 2 as those of intermediate stiffnesses and particles with swelling ratios $<$ 1.51 as stiff particles.

\subsection{Fourier transform infrared (FTIR) spectroscopy}
\begin{figure}[!t]
			\centering
			\includegraphics[width=2.7in]{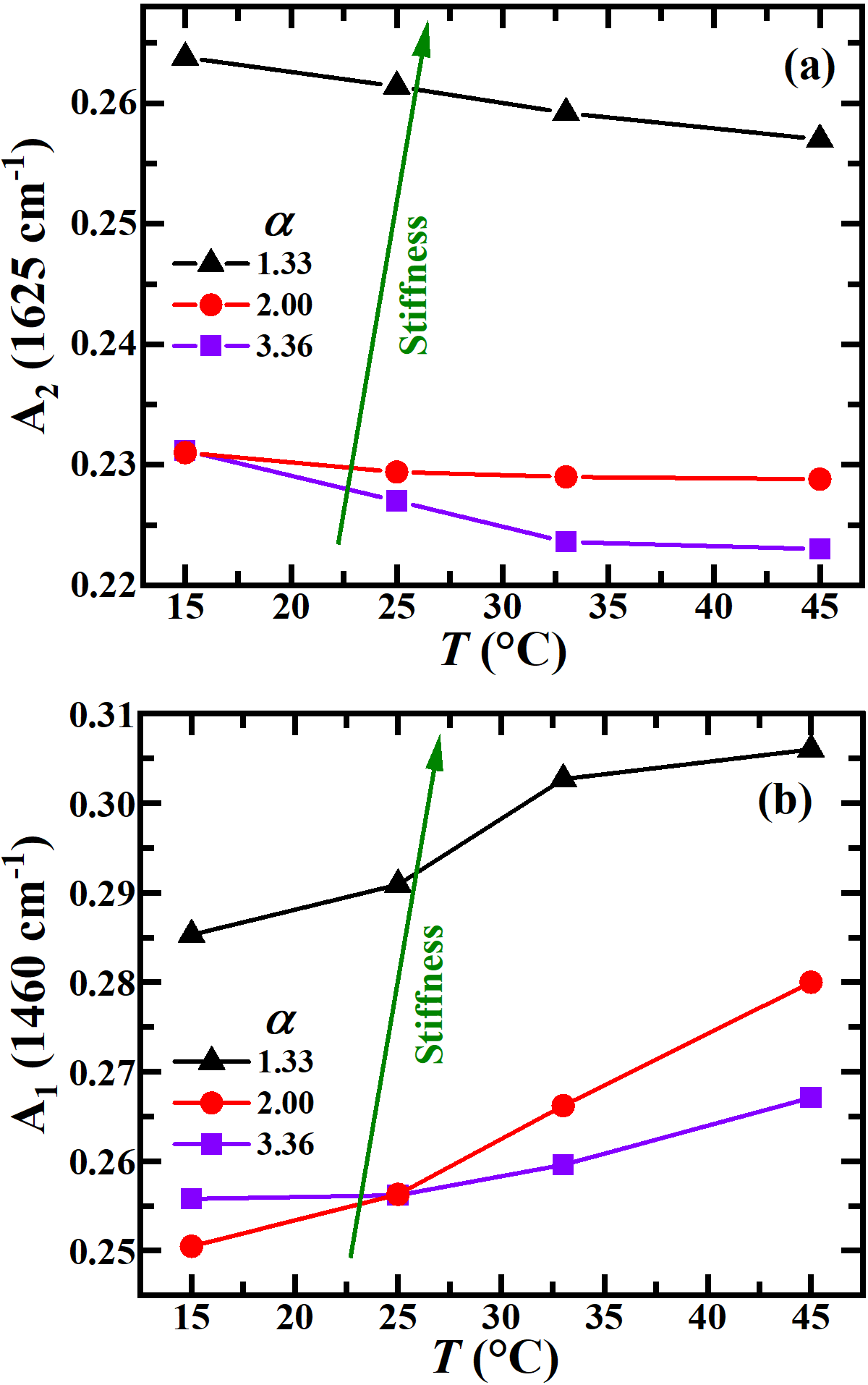}
			\caption{(a) Absorption peak heights centered at 1625 cm$^{-1}$ (obtained from FTIR spectra) {\it vs.} temperature {\it T} for suspensions constituted by PNIPAM particles of different stiffnesses, characterized by maximum swelling ratios of the particles, $\alpha$. (b) Absorption peak heights centered at 1460 cm$^{-1}$ as a function of {\it T} for aqueous suspensions of PNIPAM particles of the same stiffnesses as in (a). The spectra are displayed in Figs. S3(a-c).}	
			\label{FTIR}
		\end{figure}
FTIR spectra were measured using a Shimadzu IR Tracer-100 Fourier Transform Infrared Spectrometer. A 0.3wt.\% PNIPAM suspension prepared in D$_{2}$O was loaded in a sample cell sandwiched between two zinc sulfide (ZnS) plates separated using a teflon spacer of thickness 1 mm. The temperature of the sample cell was controlled using a water circulation unit (Julabo 300F). D$_{2}$O was used instead of H$_{2}$O to avoid overlap of the amide I band of PNIPAM (associated with \ce{C=O} stretching) with the \ce{O-H} bending mode of water \cite{He_Cheng_macromolecules_2006,Y_Maeda_Langmuir_2000}. FTIR absorption data at different temperatures are plotted in Figs. S3(a-c) of the supplementary information for PNIPAM particles of three different stiffnesses. The two absorption peaks centered around wavelengths 1625cm$^{-1}$ and 1460cm$^{-1}$ (peak heights designated respectively as A$_{2}$ and A$_{1}$ in Figs. \ref{FTIR} and S3) are respectively associated with amide I (\ce{C=O} stretching and H-bonding interactions) and amide II (\ce{N-H} bending and \ce{C-N} stretching vibrations) modes in PNIPAM \cite{He_Cheng_macromolecules_2006,Y_Maeda_Langmuir_2000}. With increase in temperature, the decrease in height of the absorption peak height A$_{2}$ in Fig. \ref{FTIR}(a) reveals the dehydration or increase in hydrophobicity of PNIPAM particles \cite{He_Cheng_macromolecules_2006,Y_Maeda_Langmuir_2000}. It can be seen from Fig. \ref{FTIR}(a) that the absorption peak height A$_{2}$ increases with increase in stiffness of PNIPAM particles, thereby clearly indicating a monotonic increase in hydrophilicity with increase in particle stiffness. This is attributed to the presence of larger crosslinker (MBA: $N,N'$-methylenebisacrylamide) concentrations of the polar groups, \ce{N-H} and \ce{C=O}<, in stiff PNIPAM particles \cite{G_Liu_2007,A_Chilkoti_2002}. The increase in absorption peak height A$_{1}$ in Fig. \ref{FTIR}(b) with increase in temperature and stiffness confirms the formation of intra- or interchain hydrogen bonds between \ce{N-H} and \ce{C=O}< \cite{He_Cheng_macromolecules_2006,Y_Maeda_Langmuir_2000}. Our FTIR data therefore demonstrates clearly that PNIPAM particle hydrophobicity decreases with increasing particle stiffness while increasing steadily with temperature.

\subsection{Zeta potential}
The zeta potentials of PNIPAM particles of different stiffnesses in aqueous suspensions were measured using an electro-acoustic accessory DT-100 purchased from Dispersion Technology \cite{A_S_Dukhin_2010}. The flat metal (gold) surface of the cylindrical electro-acoustic probe (DT-1200) was immersed in the suspension. The transducer in the probe generates ultrasound waves at a frequency of 3 MHz that propagate through the sample. The resultant oscillation of the charges in the electric double layer (EDL) around the colloidal particles produces a colloidal vibration current (CVI). The gold surface of the electro-acoustic probe detects the amplitude and phase of the CVI. The DT-1200 software is used to analyze the CVI data and compute zeta potentials. More details of the instrument and electro-acoustic principles can be found in the literature \cite{A_S_Dukhin_2010}. To check the accuracy of our setup, we measured the zeta potential of a standard sample, 10wt.\% of Silica Ludox prepared in 0.01M KCl with an expected standard value of -38 mV, and noted a readout of -37.5 $\pm$ 1.12 mV at a temperature $25^{\circ}$C. 
\begin{figure}[!t]
			\centering
			\includegraphics[width=3.5in]{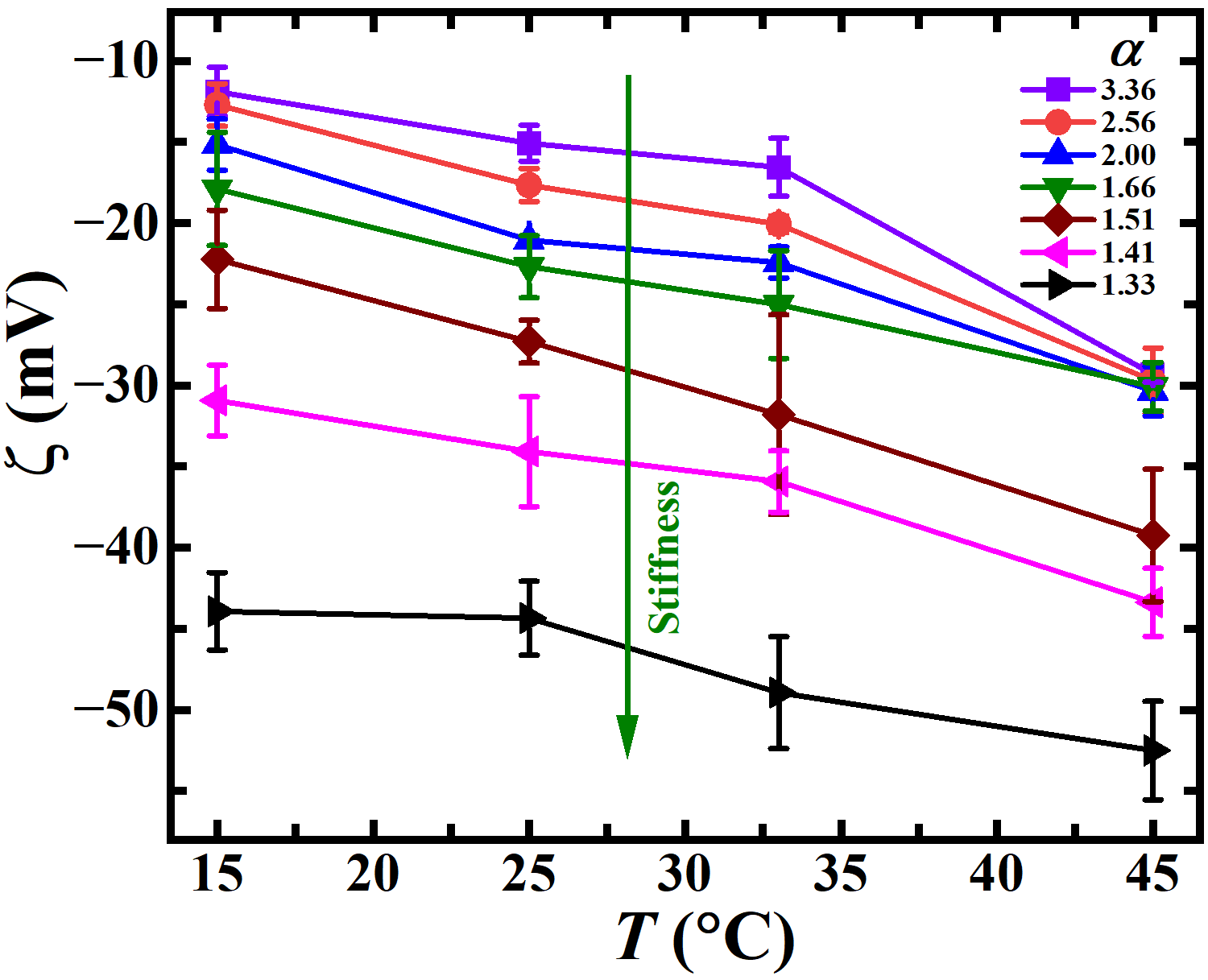}
			\caption{Temperature-dependent zeta potentials of 1wt.\% aqueous suspensions of PNIPAM particles of different maximum swelling ratios. Error bars are the standard deviations estimated from four independent measurements. The direction of the  arrow indicates increase in particle stiffness.}	
			\label{Zeta potential}
		\end{figure}
\paragraph{}
At temperatures below the LCST, PNIPAM particles are stabilized by a combination of steric repulsion and electrostatic interactions \cite{T_Adam_2019}. On the other hand, at temperatures above the LCST, PNIPAM particles in aqueous suspension are expected to be hydrophobic in nature. Fig. \ref{Zeta potential} shows the plot of the zeta potential of PNIPAM particles with different maximum swelling ratios, $\alpha$, as a function of temperature. It has been reported that the minimum electrostatic potential required for stabilizing colloidal particles against aggregation is $\approx$ $\pm$30 mV \cite{hunter2013zeta}. From Fig. \ref{Zeta potential}, we see that the magnitudes of zeta potential increase with increasing stiffness and temperature. Stiffer particles in aqueous suspension are therefore always repulsive and significantly more stable when compared to particles that are soft or of intermediate stiffnesses. This can be attributed to the increase in ionic groups originating from the polymerization of the crosslinker MBA during synthesis, which effectively stabilizes PNIPAM particles against aggregation. We note from Fig. \ref{Zeta potential} that even at temperatures below the LCST, zeta potentials of PNIPAM particles that are soft or of intermediate stiffnesses are not high enough to resist aggregation in aqueous suspension. Contrary to expectation, these PNIPAM particles may therefore aggregate even below the LCST. Our results therefore reveal that soft PNIPAM particles are electrostatically unstable at all temperatures in the range explored here. We note that our zeta potential values are in good agreement with the zeta potential measurements of PNIPAM particles in suspension that were reported by Town {\it et. al.} \cite{T_Adam_2019}. From the FTIR data discussed in section 2.3, we noted that the hydrophobicity of PNIPAM particles in aqueous suspensions increases with increasing temperature and with decreasing particle stiffness. PNIPAM particle self-assembly can therefore be expected to be very sensitive to  the competition between inter-particle hydrophobic and electrostatic interactions.

\subsection{Rheology}
Rheological measurements were performed using stress-controlled Anton Paar MCR 501 and MCR 702 rheometers. A double gap geometry with a gap of 1.886 mm, an effective length of 40 mm and requiring a sample volume of 3.8 ml was used for dilute suspensions. For dense suspensions, a cone-plate geometry with a cone radius $r_{c}$ = 12.491 mm, cone angle of 0.979$^{\circ}$, measuring gap $d$ = 0.048 mm and requiring a sample volume of 0.07 ml was used. A Peltier unit and a water circulation system (Viscotherm VT2) capable of controlling temperatures between $0^{\circ}$C - $180^{\circ}$C and $5^{\circ}$C - $80^{\circ}$C were respectively used with cone-plate and double gap sample geometries. Silicone oil of viscosity 5 cSt was used as a solvent trap oil to avoid water evaporation. The viscoelastic properties of dense aqueous suspensions of PNIPAM, constituted by particles of different stiffnesses, were investigated at different suspension temperatures by performing oscillatory rheological experiments. For oscillatory frequency sweep measurements, the peak-to-peak strain amplitude was kept fixed at 0.5\% while ramping up/ down the applied angular frequencies logarithmically. For oscillatory amplitude sweep measurements, the applied angular frequency was kept constant at 0.5 rad/sec and the applied strain amplitudes were increased logarithmically.
\begin{figure}[!t]
			\centering
			\includegraphics[width=3.5in]{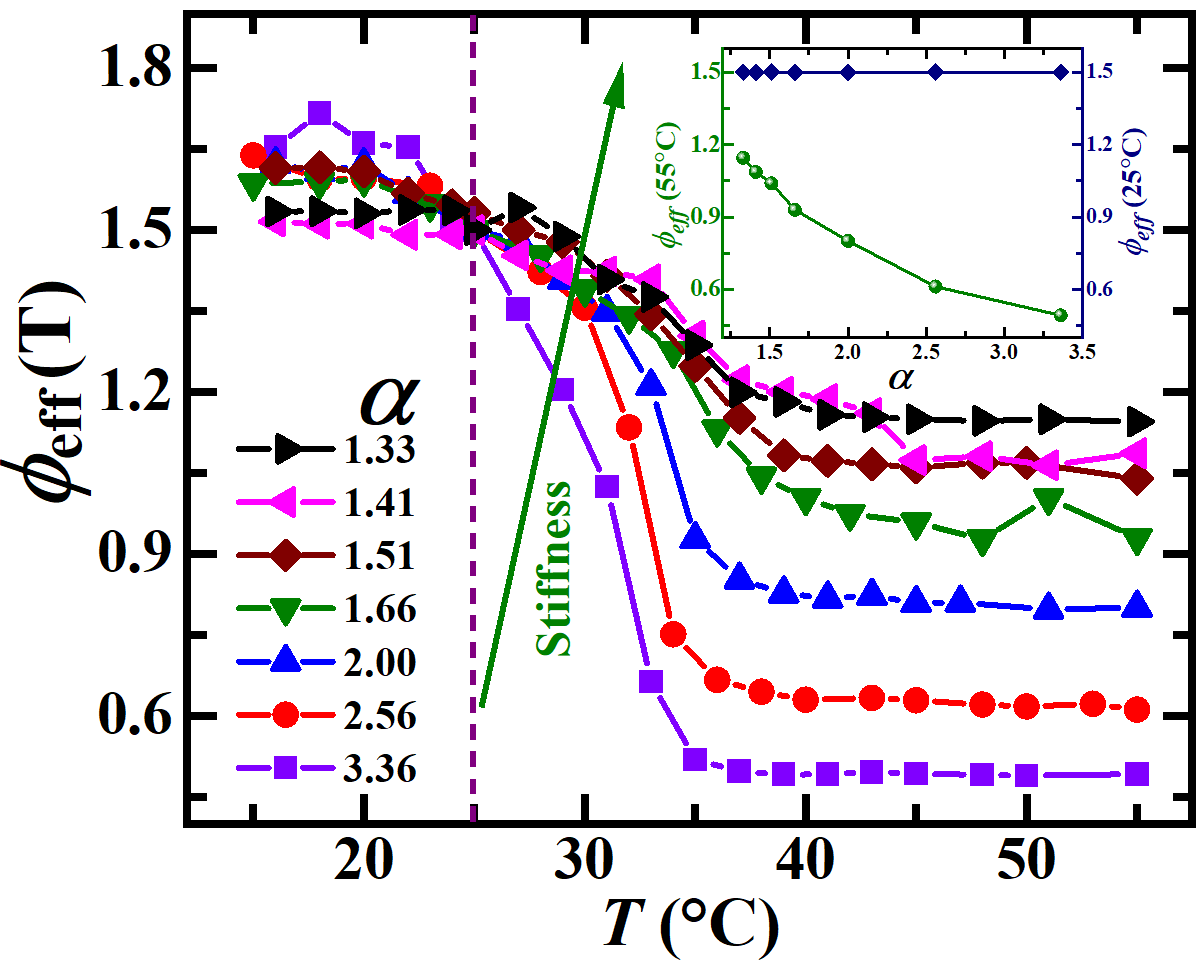}
			\caption{Temperature-dependent $\phi_{eff}$ of aqueous suspensions of PNIPAM particles of different maximum swelling ratios, $\alpha$. Samples are prepared so that all suspensions have an effective volume fraction of 1.5 at 25$^{\circ}$C (indicated by the vertical purple dashed line). The olive arrow points towards the direction of increasing particle stiffness, which coincides with the direction of decrease in maximum swelling ratio, $\alpha$. The inset shows effective volume fractions of suspensions constituted by particles of varying stiffnesses (or swelling ratios, $\alpha$) at temperatures 55$^{\circ}$C (above LCST) and 25$^{\circ}$C (below LCST). While all the suspensions have $\phi_{eff}$ = 1.5 below the LCST, $\phi_{eff}$ varies inversely with $\alpha$ above the LCST because of the significantly faster deswelling rates of particles of lower stiffnesses.}	
			\label{Phi(T)}
		\end{figure}
\subsection{Preparation of dense aqueous PNIPAM microgel suspensions}
Dense aqueous PNIPAM suspensions were prepared by adding PNIPAM powder in Milli-Q water. The samples were stirred for 24 h and sonicated for 15 min. Dense suspensions were diluted to prepare PNIPAM suspensions of lower concentrations. Since PNIPAM particles are deformable, volume fraction $\phi$ is not an appropriate measure to quantify their packing in aqueous suspensions. Therefore, a modified parameter called effective volume fraction, $\phi_{eff}$, is used here. The protocol for estimating the effective volume fraction of PNIPAM particles in suspension, synthesized by varying crosslinker concentrations during the synthesis process, is discussed in detail in section 3 of the supplementary information. In the present work, the concentrations of ingredients were chosen to ensure an effective volume fraction $\phi_{eff}$ = 1.5 at $T$ = 25$^{\circ}$C. We have estimated the temperature-dependent $\phi_{eff}$ using the relation $\phi_{eff}(T) = \phi_{eff}(25^{\circ}C)\times(\frac{d_{h}(T)}{d_{h}(25^{\circ}C)})^3$ \cite{V_Carrier_JOL_2009}, where $\phi_{eff}(T)$, $\phi_{eff}(25^{\circ}C)$, $d_{h}(T)$ and $d_{h}(25^{\circ}C)$ are $\phi_{eff}$ at temperatures $T$ and $25^{\circ}$C and average hydrodynamic diameters of PNIPAM particles at $T$ and $25^{\circ}$C, respectively. The temperature-dependent $\phi_{eff}(T)$ of aqueous suspensions of PNIPAM particles ($\phi_{eff}$ = 1.5 at $25^{\circ}$C) synthesized with different crosslinker concentrations are plotted in Fig. \ref{Phi(T)}. Even though $\phi_{eff}$ is the same for all suspensions at 25$^\circ$C (which lies below the LCST),  we note from the inset of Fig. \ref{Phi(T)} that the effective volume fraction, $\phi_{eff}$, decreases rapidly with increase in maximum swelling ratio, $\alpha$, at a temperature above the LCST. This is because softer particles deswell much more rapidly than stiffer particles. 
		
\subsection{Differential scanning calorimetry}
Differential scanning calorimetry (DSC) (Mettler Toledo, DSC 3) measurements were performed to estimate the LCSTs of PNIPAM particles of different maximum swelling ratios, $\alpha$, in dense aqueous suspensions. The heat flows evaluated in the DSC experiments for aqueous suspensions of PNIPAM particles of different $\alpha$ are plotted in Fig. S5(a) of the supplementary information as a function of temperature $T$. The temperatures corresponding to the endothermic peaks are characterized as the LCSTs of the PNIPAM particles. Fig. S5(b) shows the LCST values of suspensions of PNIPAM particles of different $\alpha$. We note a slight decrease in LCST of PNIPAM particles in aqueous suspension with increase in $\alpha$. The stiffer particles are more hydrophilic than the softer particles due to the presence of a larger number of polar groups (N-H and C=O) arising from the higher concentrations of the crosslinker ($N,N'$-methylenebisacrylamide) used during their synthesis. The LCSTs of stiffer particles are therefore higher.

\subsection{Cryogenic scanning electron microscopy (cryo-SEM)}
The microstructures of dense aqueous suspensions of PNIPAM particles of different maximum swelling ratios, prepared at temperatures below, near and above the LCST, were studied using cryogenic scanning electron microscopy (cryo-SEM). The suspensions were loaded in a ribett fixed with a cryo-stub and then cryo-frozen at -$190$$^{\circ}$C by plunging into nitrogen slush at atmospheric pressure. The frozen sample was loaded into the preparation chamber with a cryo-transfer holder. The sample was sublimated for $5$ min at a temperature of -$90$$^{\circ}$C and then fractured with a stainless steel knife. The fractured sample was again sublimated for $15$ min at -$90$$^{\circ}$C to remove the ice formed over the surface and then coated with a thin layer of platinum in vacuum conditions using a cryo-transfer system (PP3000T from Quorum Technologies). Images of the samples were reconstructed by collecting backscattered electrons obtained with a field-effect scanning electron microscope (FESEM), Ultra Plus FESEM-4098 from Carl Zeiss, at an electron beam strength of 5 KeV.
		
\section{Results and Discussion}
\subsection{Temperature-dependent swelling and mechanical properties of PNIPAM suspensions }
We systematically study the temperature-dependence of the self-assembled structures, rheology and phase behavior of dense aqueous suspensions of PNIPAM particles of different stiffnesses. The temperature-dependent mechanical properties of these suspensions are estimated by performing temperature sweep rheological experiments at a fixed oscillatory strain amplitude $0.5\%$ and angular frequency 1 rads$^{-1}$. Fig. \ref{Temp_test}(a) shows a plot of the elastic and viscous moduli, $G'$ and $G''$, as a function of temperature $T$ for aqueous suspensions of PNIPAM particles of different stiffnesses, having effective volume fraction $\phi_{eff}$ = 1.5 at 25$^{\circ}$C. Below the LCST, we observe that both $G'$ and $G''$ increase with increase in particle stiffness. This is in agreement with previous work that reported an increase in suspension elasticity with increase in particle crosslinker density \cite{Senff_H_1999}. According to Mattsson $et$ $al.$ \cite{J_Mattsson_2009}, the fragility of a colloidal glass is determined by particle-scale elasticity. Since stiff particles swell less than soft particles in aqueous suspension, we used a larger concentration of the former to achieve the same effective volume fraction at 25$^{\circ}$C (Table 1). Furthermore, it has been shown {\it via} osmotic pressure measurements that the shear modulus of soft microgel particles is much smaller when compared to their bulk modulus, implying that the particles are more likely to deform rather than compress in dense packings \cite{A_scotti_PRE_2021}. We expect that the contribution of shear modulus becomes less significant with increase in crosslinker concentration, thereby altering the particle-scale packing behavior. The larger $G'$ values of suspensions of stiffer particles (observed in Fig. \ref{Temp_test}) can be attributed to the increase in particle elasticity in the presence of higher crosslinker concentration and higher particle number density. For all the samples, $G'$ > $G''$, at $T<$ LCST, indicating viscoelastic solid-like behavior at these temperatures due to kinetic constraints imposed upon the swollen PNIPAM particles. As expected, we observe that both moduli decrease sharply as the LCST is approached (Figs. \ref{Temp_test}(a-b)) due to reduction in the sizes of the constituent particles (Fig. \ref{DLS}). We also note that deswelling of PNIPAM particles near the LCST results in a significant decrease in the effective particle packing fraction, as seen earlier in  Fig. \ref{Phi(T)}. Both viscoelastic moduli are minimum at their LCSTs. We see from Fig. S6 of the supplementary information that the LCSTs estimated from the rheology and DSC data show excellent overlap. On raising the temperature beyond the LCST, we observe a gradual increase in the viscoelastic moduli, with $G'$ eventually exceeding $G''$ for all the suspensions.
\begin{figure}[!t]
\centering
\includegraphics[width=2.8in]{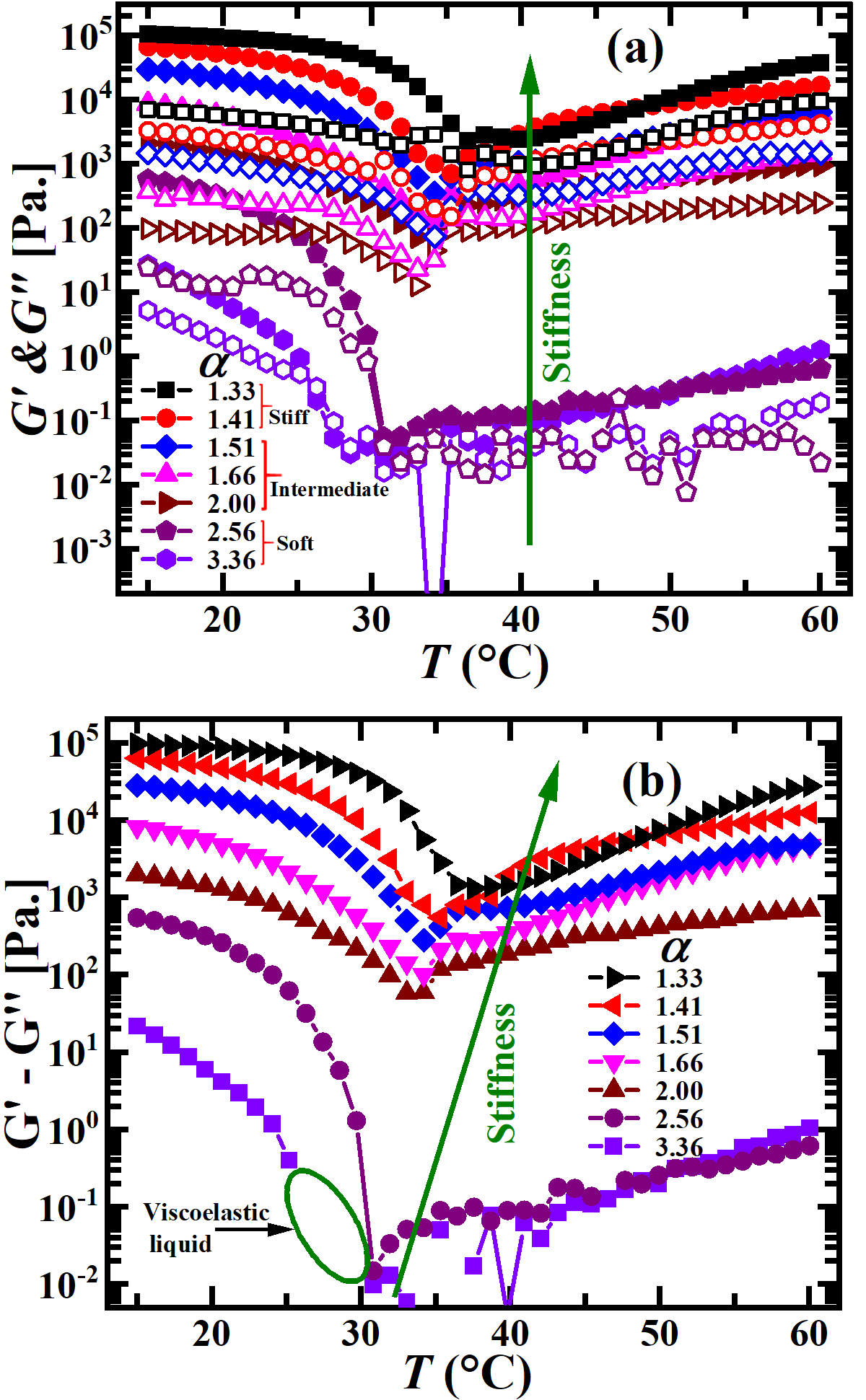}
\caption{(a) Temperature-dependent elastic moduli, $G'$ (solid symbols) and viscous moduli $G''$ (hollow symbols), of aqueous suspensions of PNIPAM particles of different stiffnesses (or maximum swelling ratios, $\alpha$), prepared at effective volume fraction, $\phi_{eff}$ = 1.5, at 25$^{\circ}$C. (b) $G'$ - $G''$ $vs.$ temperature $T$ for aqueous suspensions of PNIPAM particles of different maximum swelling ratios, $\alpha$. }	
\label{Temp_test}
\end{figure}
\begin{figure*}[!t]
\centering
\includegraphics[width=6.8in]{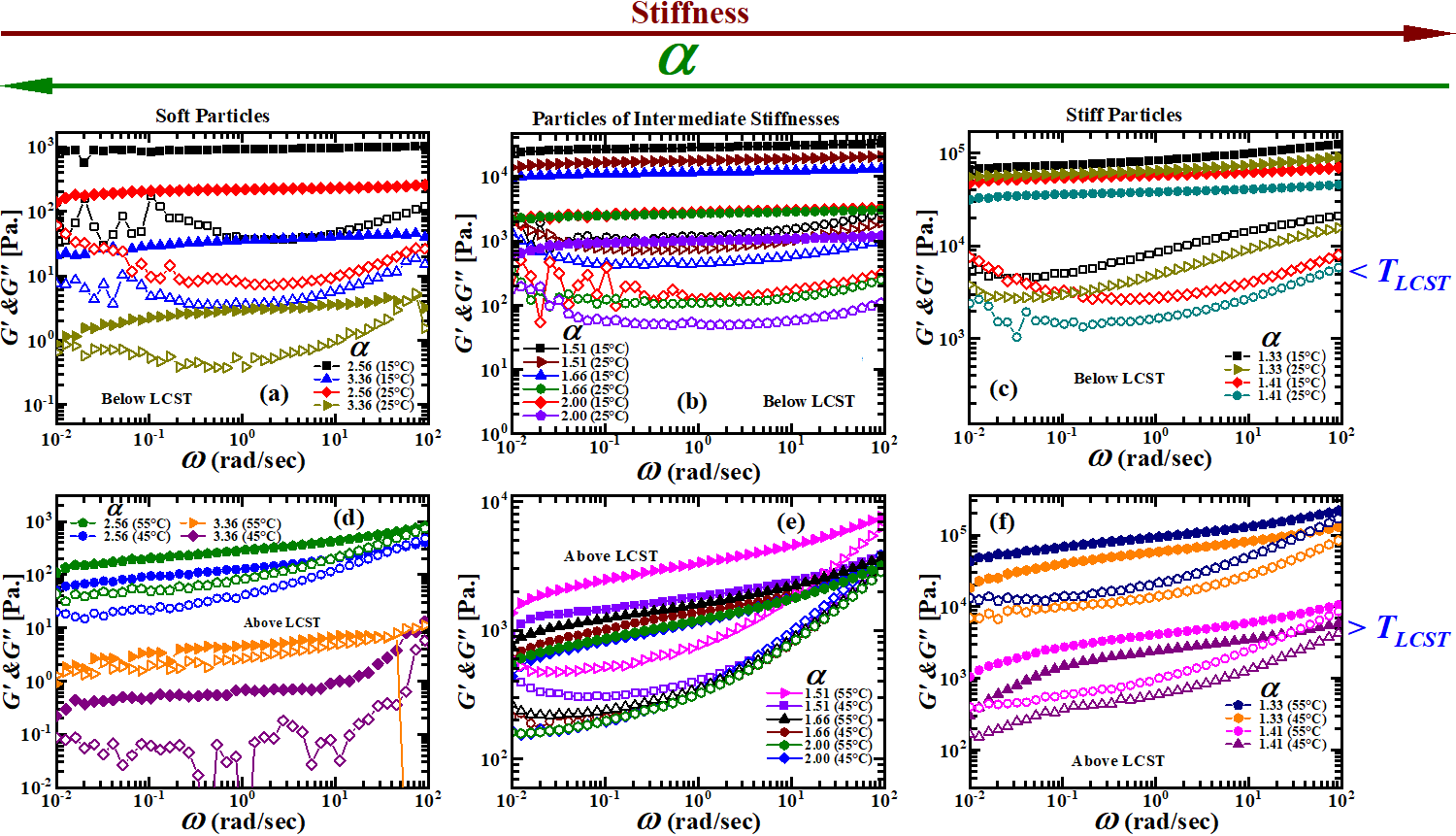}
\caption{Elastic moduli, $G'$ (solid symbol), and viscous moduli, $G''$ (hollow symbols), as a function of applied oscillatory frequency, $\omega$, for aqueous PNIPAM suspensions prepared at $\phi_{eff}$ = 1.5 at 25$^{\circ}$C and constituted by soft particles (left panel), particles of intermediate stiffnesses (middle panel) and stiff PNIPAM particles (right panel) at temperatures (a-c) below the LCST and (d-f) above the LCST.}
\label{FS_all}
\end{figure*}
\paragraph{}
Fig. \ref{Temp_test}(b) displays plots of $G'$ - $G''$ over the entire range of temperatures explored here. For all the suspensions, we observe a significant loss of rigidity near the LCST. Interestingly, we note that $G''$ > $G'$ near the LCST only for the suspension of the softest PNIPAM particles ($\alpha$ = 3.36, indicated by the olive circle in Fig. \ref{Temp_test}(b)), revealing liquid-like behavior. Using suspensions of PNIPAM particles of a fixed swelling ratio, Romeo $et$ $al.$ \cite{Romeo2010} had reported a transformation from a repulsive glass to an attractive gel phase while passing through a liquid-like state near the LCST. The increase in suspension rigidity at temperatures above the LCST can be attributed to {the space-spanning particle assemblies that are expected to form due to the enhanced attractive interactions between the hydrophobic PNIPAM particles in aqueous suspension. We see from our zeta potential measurements in Fig. \ref{Zeta potential} that stiff PNIPAM particles are electrostatically stable at temperatures above the LCST. Since our FTIR data clearly indicates a significant enhancement of hydrophobicity with temperature, we conclude that the observed increase in viscoelastic moduli above the LCST results from strong hydrophobic attractions that dominate over the electrostatic repulsion at high temperatures. Interestingly, we observe that aqueous suspensions of stiff particles do not exhibit a liquid phase near the LCST. Stiffer particles are characterized by increased particle elasticity and show reduced de-swelling when compared to soft particles. These factors, along with the} comparatively higher effective particle volume fractions of stiff particle suspensions can explain the absence of a liquid phase at the LCST in these suspensions. Clearly, the solid-solid transition of dense aqueous suspensions of PNIPAM particles across the LCST is strongly dependent on the stiffness of the constituent particles.

\subsection{ Influence of underlying suspension microstructures on macroscopic mechanical properties} 

\subsubsection{Frequency sweep rheology to explore linear rheological properties}

We next perform oscillatory frequency sweep rheological experiments at a fixed strain amplitude, $\gamma$ = 0.5\%, to study the dynamics of dense aqueous suspensions of PNIPAM particles of different stiffnesses at temperatures below, near and above the LCST. Fig. \ref{FS_all} shows oscillatory frequency sweep data of PNIPAM particles of different stiffnesses performed at temperatures below and above the LCST (top and bottom panels respectively). Frequency sweep data taken at temperature near the LCST are shown in Fig. S7 of the supplementary information. Our frequency sweep data reveal the presence of enhanced mechanical moduli in suspensions constituted by stiffer particles. This is in agreement with the temperature sweep rheological data plotted in Fig. \ref{Temp_test}(a). Below the LCST, all PNIPAM suspensions display weak or no frequency-dependence of $G'$ and weakly frequency-dependent $G''$ (Figs. \ref{FS_all}(a-c)), that indicate the existence of slow glassy dynamics in these samples. For the suspension constituted by the softest PNIPAM particles ($\alpha$ = 3.36, Fig. \ref{FS_all}(a)), we observe the existence of two crossover points in the high and low-frequency regimes, presumably arising from two different ($\alpha$ and $\beta$) relaxation mechanisms. Crossover points for the other samples are expected to occur outside the frequency window explored in this study. The noisy frequency sweep data for aqueous suspensions of soft PNIPAM particles at a temperature near the LCST (Fig. S7(a)) arise due to the very low mechanical moduli of the liquid-like samples under these conditions. In contrast, the frequency-dependent viscoelastic behavior of suspensions of PNIPAM particles of intermediate and large stiffnesses near the LCST (Figs. S7(b) and (c)) indicates the presence of weak gels/ glasses. Above the LCST, the weak increases of both moduli with applied frequency (Figs. \ref{FS_all}(d-f)) are typical characteristics of glasses or gels composed of hard attractive colloids \cite{K_N_pham_JOL_2008,V_Prasad_2003,V_Trappe_2000}. This observation indicates the formation of gel/ glass-like phases formed by the collapsed PNIPAM particles and driven by attractive hydrophobic interactions that are expected to dominate under these conditions (Fig. \ref{FTIR}). We note that the mechanical moduli of the suspensions comprising soft particles are considerably lower above the LCST than below it. Moduli characterizing the suspensions of stiffer particles below and above the LCST, in contrast, have comparable magnitudes. 

\subsubsection{Amplitude sweep rheology to explore yielding properties}
\begin{figure*}[!t]
\centering
\includegraphics[width=6.8in]{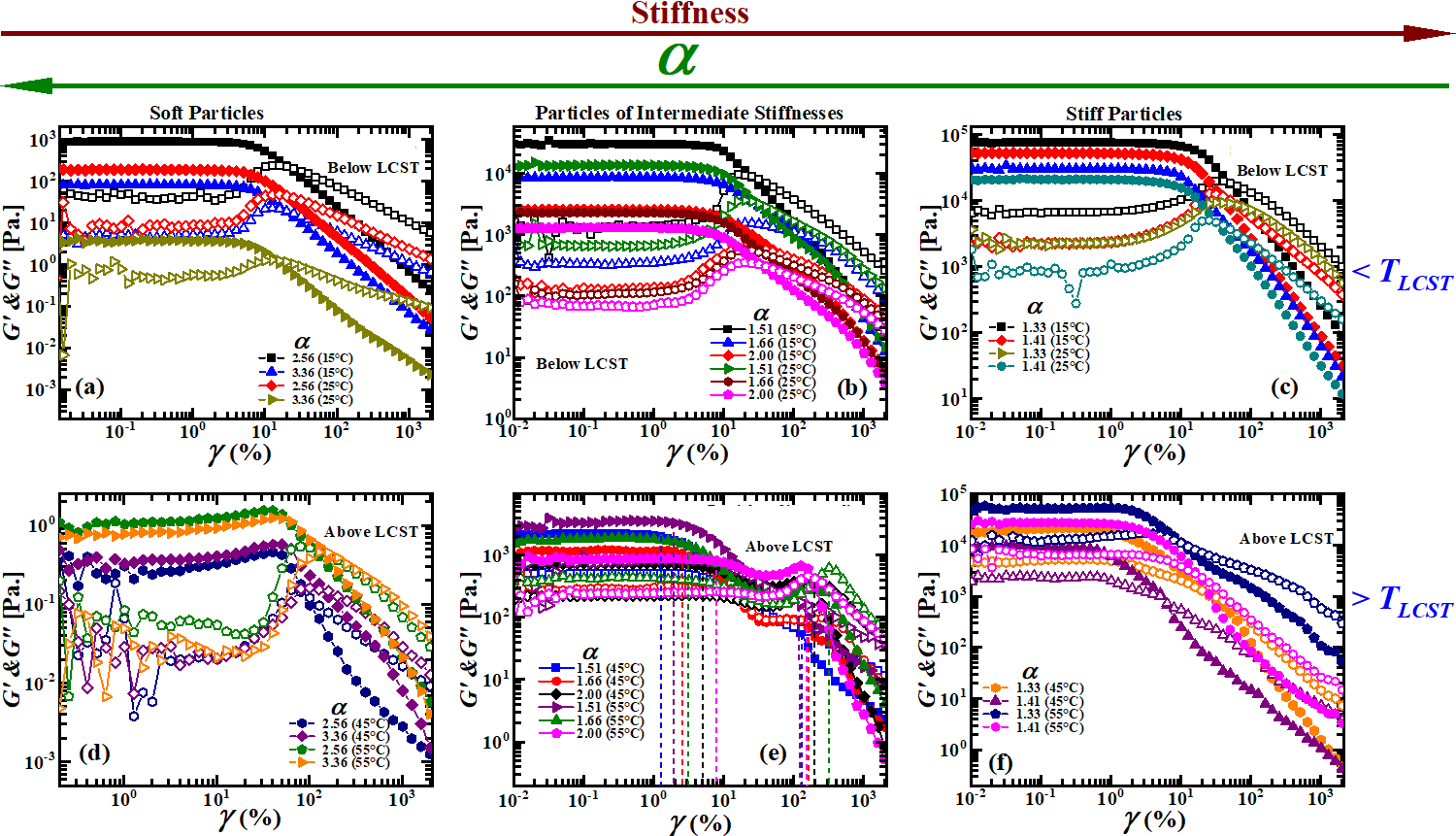}
\caption{Elastic moduli, $G'$ (solid symbol), and viscous moduli, $G''$ (hollow symbols), as a function of applied strain amplitude, $\gamma$, for aqueous PNIPAM suspensions, prepared at $\phi_{eff}$ = 1.5 at 25$^{\circ}$C and constituted by soft particles (left panel), particles of intermediate stiffnesses (middle panel) and stiff PNIPAM particles (right panel) at temperatures (a-c) below the LCST and (d-f) above the LCST.}	
\label{AS_all}
\end{figure*}
We perform oscillatory strain amplitude sweep rheological experiments at a fixed angular frequency $\omega$ = 0.5 rads$^{-1}$ to study the yielding behaviors of dense aqueous PNIPAM suspensions comprising PNIPAM particles of different stiffnesses and at various suspension temperatures. Data acquired from all samples at temperatures below and above the LCST are plotted in Fig. \ref{AS_all}, and data at temperatures near the LCST are plotted in Fig. S8 of the supplementary information. For all samples below the LCST (Fig. \ref{AS_all} (a-c)), $G'$ > $G''$ with both $G'$ and $G''$ seen to be independent of the applied strain amplitude for small strain amplitudes. The samples therefore all exhibit viscoelastic solid-like behavior at small strains. With increase in applied strain amplitude, the suspensions start yielding, as indicated by a monotonic decrease in the elastic modulus, $G'$, at higher strains. Simultaneously, the viscous modulus, $G''$, shows a peak before decreasing, which signals the onset of an yielding process. $G''$ eventually exceeds $G'$ at the highest strains indicating a fluid-like mechanical response. While the magnitudes of the moduli differ, we note that these trends, regarded as typical features of soft glassy materials \cite{Mason_PRL_1995,Peter_PRE_1998}, are visible in all samples below and above the LCST. The yield stresses and strains are calculated from amplitude sweep data following the method proposed by Laurati $et$ $al.$ \cite{M_Laurati_2011}. The details of the analysis and a representative plot are presented in section 5.1 and Fig. S9 of the supplementary information. In Figs. S10(a) and (b), we plot the temperature-dependent linear moduli, $G_{l}'$, defined as the magnitude of $G'$ at very low applied strain amplitudes, and first yield stresses, $\sigma_{y}$, respectively, for all the suspensions investigated here. We see from Fig. S10(a) that $G_{l}'$ increases with particle stiffness. From Fig. S10(b), we note that $\sigma_{y}$ values are the lowest for suspensions of soft particles at temperatures near the LCST. Suspensions of soft PNIPAM particles therefore shear melt at smaller stresses and are mechanically very fragile. First yield strains are also calculated by adopting the method suggested in \cite{M_Laurati_2011} and plotted in Fig. S11 of the supplementary information. While yield strains below the LCST are of comparable magnitudes, we note that at temperatures above the LCST, yield strains of suspensions of soft PNIPAM particles are much higher than those constituted by particles of high and intermediate stiffnesses. The observed increase in yield strains of suspensions constituted by soft particles can be attributed to their lower effective volume fraction and mechanical fragility, wherein higher structural and bonding flexibility allow stretching and elongation before bond breaking at higher strains \cite{N_Koumakis_2011}.
\begin{figure*}[!t]
\centering
\includegraphics[width=6in]{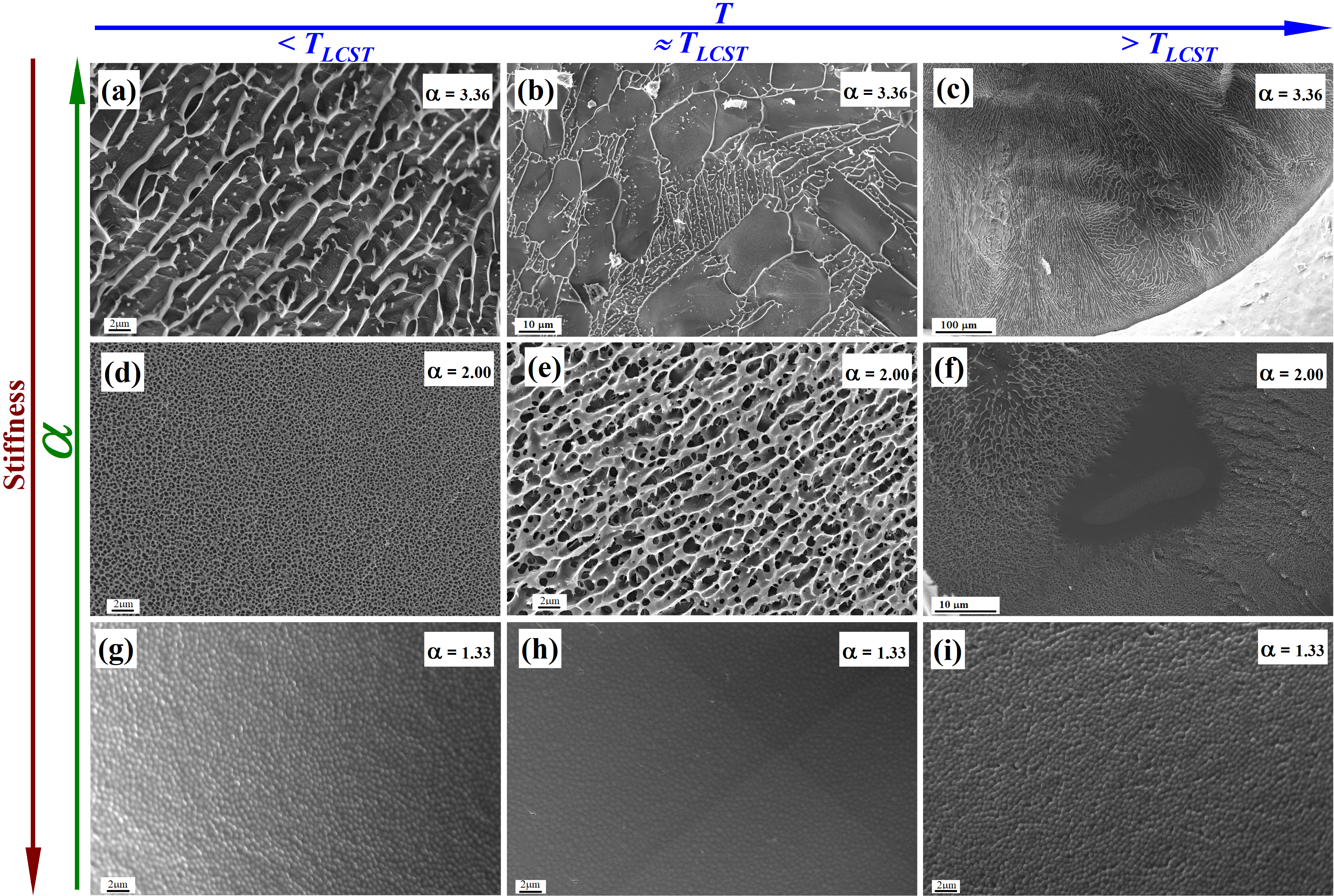}
\caption{Cryo-SEM micrographs of aqueous suspensions of (a-c) soft PNIPAM particles, (d-f) PNIPAM particles of intermediate stiffnesses and (g-i) stiff PNIPAM particles at temperatures below the LCST (left panel), near the LCST (middle panel) and above the LCST (right panel). The samples are prepared at $\phi_{eff}$ = 1.5 at 25$^{\circ}$C, stabilized at the pre-determined temperature, and then frozen to cryogenic temperatures.}
\label{Cryo-sem}
\end{figure*}
\paragraph{}
Interestingly, dense suspensions of PNIPAM particles of intermediate stiffnesses display two-step yielding above the LCST (Fig. \ref{AS_all}(e); indicated by vertical dashed lines), resembling the yielding of glasses/gels comprising hard attractive colloidal particles\cite{K_N_pham_JOL_2008}. The first yield point ($\gamma_{y}$) can be attributed to inter-cluster bond breaking of the caged particles constituting the dense suspensions, while the second yield point ($\gamma_{y2}$) likely arises from the breaking of cages or clusters into smaller fragments \cite{N_Koumakis_2011,Zhou_2014_lang}. The absence of two-step yielding in suspensions of soft PNIPAM particles above the LCST can be attributed to their mechanical fragility and relatively lower effective volume fractions. For suspensions of stiffer PNIPAM particles, the highly crowded environment ensures that bond and cage-breaking phenomena coexist simultaneously, resulting in one-step yielding of these suspensions under high strains (Fig. \ref{AS_all}(f)) \cite{N_Koumakis_2011,Zhou_2014_lang}. Our strain amplitude sweep data therefore verify our earlier conclusions reported in section 3.2.1 that suspensions of stiff PNIPAM particles synthesized using higher crosslinker concentrations exhibit characteristics of attractive hard colloidal glasses/ gels, while suspensions of soft particles form relatively weaker gels below and above the LCST.

\subsubsection{Cryogenic scanning electron microscopy (cryo-SEM) to directly visualize suspension microstructures}

One of the key motivations of this study was to probe the mechanical properties of solid-like phases of PNIPAM suspensions constituted by particles of various stiffnesses over a broad range of temperatures across the LCST. We now attempt to understand our rheology data in terms of the underlying self-assembled colloidal suspension microstructures via direct visualization using a cryogenic scanning electron microscope (cryo-SEM). In Fig. \ref{Cryo-sem}, we show the temperature and stiffness-induced morphological changes in dense aqueous suspensions of soft PNIPAM particles ($\alpha$ = 3.36; Fig. \ref{Cryo-sem}(a-c)), particles of intermediate stiffnesses ($\alpha$ = 2.00; Fig. \ref{Cryo-sem}(d-f)) and stiff particles ($\alpha$ = 1.33; Fig. \ref{Cryo-sem}(g-i)), prepared at temperatures below the LCST ($15^{\circ}$C, left panel), near the LCST ($33^{\circ}$C, middle panel) and above the LCST ($55^{\circ}$C, right panel). Additional cryo-SEM micrographs of suspensions of PNIPAM particles of maximum swelling ratios $\alpha$ = 2.56, 1.51 and 1.41 are shown in Fig. S12 of the supplementary information. Dense aqueous suspensions of the relatively softer PNIPAM particles display three-dimensional network structures (Figs. \ref{Cryo-sem}(a-f) and S12(a-f)). Magnified cryo-SEM images of PNIPAM particles of two different maximum swelling ratios at a temperature below the LCST have been shown in Fig. S13 where individual PNIPAM particles can be clearly seen. We believe that the observed networks comprise strands of PNIPAM particles connected to form system-spanning structures displaying gel-like rheology as also reported in earlier studies \cite{M_Laurati_2011,Erramreddy_2017,Nair_2019}. In Fig. 6(a), we had reported two characteristic timescales in the frequency response of suspensions of soft particles below the LCST. These timescales are presumably set by the motion of single particles on the one hand, and the dynamics of the gel network strands on the other \cite{N_Koumakis_2011}. 
Cryo-SEM micrographs of these suspensions clearly reveal the existence of inhomogeneous networks at temperatures near and above the LCST (Figs. \ref{Cryo-sem}(b-c) and S12(b-c)). Above the LCST, the images (Figs. \ref{Cryo-sem}(c) and (f)) show the existence of phase separation that presumably arises due to enhanced hydrophobic interactions between PNIPAM particles. In comparison, the micrographs for the suspensions constituted by stiffer particles in Figs. 8(g-i) and S12(g-i) show dense disordered packings of particles without any clearly distinguishable interconnected clusters at temperatures both below and above the LCST.

\paragraph{}
Our zeta potential measurements plotted in Fig. \ref{Zeta potential} revealed that softer particles are electrostatically unstable due to weak surface potentials and can therefore aggregate even below the LCST. Hydrophobic attractions are also stronger between the hydrophobic moieties constituting softer particles when compared to stiffer ones as discussed in section 2.3. The combination of these two factors and the strong temperature-dependent enhancement of hydrophobicity in suspensions of soft particles and particles of intermediate stiffnesses give rise to attractive gel-like structures both above and below the LCST, as observed in Figs. \ref{Cryo-sem}(a-f) and S12(a-f). We adopt a protocol used earlier \cite{Samim_2016} to estimate the average areas of the porous microstructures displayed in the aforementioned figures and display the data in Fig. S14.
\paragraph{}
The number density of particles is highest in suspensions of stiff particles as highlighted earlier in section 2.6 and Table 1. For suspensions of the stiffest particles, therefore, the networks formed are extremely dense and the suspensions remain in a glassy state at temperatures below and above the LCST. In these samples, electrostatic repulsion dominates due to the abundance of charged species used to polymerize the crosslinker during particle synthesis. As a result, the particles assemble to form dense disordered packings reminiscent of repulsive glasses at a temperature below the LCST (Fig. \ref{Cryo-sem}(g) and Fig. S12(g)). On the other hand, at temperatures above the LCST where hydrophobic attractions dominate over the electrostatic repulsion as revealed by our rheological measurements, we observe attractive glassy phases (Fig. \ref{Cryo-sem}(i) and Fig. S12(i)). The close-packed structures formed in these suspensions give rise to higher mechanical moduli as discussed in sections 3.2.1 and 3.2.2. Our work clearly reveals that the self-assembly of dense aqueous suspensions of PNIPAM particles depends on the suspension temperature and stiffness of the constituent particles, and can be explained by studying the competition between inter-particle hydrophobic and electrostatic interactions. 

\section{Conclusions}
The present study demonstrates that the mechanical properties and self-assembled morphologies of dense PNIPAM suspensions can be tuned by controlling the suspension temperature and particle stiffness. In this work, we synthesized PNIPAM particles of different stiffnesses by varying the crosslinker concentration in a free radical precipitation polymerization method. We quantified the stiffnesses of the synthesized particles by measuring their maximum swelling ratios. Our dynamic light scattering data reveal that the soft particles, characterized by higher osmotic pressures, deswell more rapidly than stiff particles upon increasing temperature. We performed Fourier transform infrared spectroscopy experiments to study the temperature-dependent hydrophobicity of aqueous PNIPAM suspensions. Analysis of the data shows that the hydrophobicity of PNIPAM particles increases with temperature and decreases with particle stiffness. Since the concentration of the polar crosslinker, MBA, used during synthesis of stiff PNIPAM particles was much higher than the amount required to synthesize the softer ones, stiff particle suspensions are characterized by a larger number density of ionic groups and therefore enhanced hydrophilicity. To study the competition between hydrophobic and electrostatic interactions, we estimated surface charge densities of PNIPAM particles of different stiffnesses in aqueous suspensions by measuring surface zeta potentials. Interestingly, we see clear signatures of the presence of electrostatic attraction between soft PNIPAM particles and those of intermediate stiffnesses even at temperatures below the LSCT. On the other hand, suspensions constituted by stiff PNIPAM particles are stable against electrostatic attraction over the entire temperature range.\\ 

Furthermore, we observe a non-monotonic increase in suspension viscoelasticity when temperature is increased across the LCST. We note that the suspensions constituted by the softest PNIPAM particles exhibit a gel-liquid-gel transition with increase in temperature, with the gel phase above the LCST being significantly weaker than that formed below the LCST. On the other hand, PNIPAM microgels constituted by stiff particles form disordered glassy phases below and above the LCST, while retaining finite though low rigidities even at the LCST. Interestingly, in an earlier study, a similar increase in suspension rigidity above the LCST was attributed to a change in interparticle interaction from repulsive to attractive due to enhancement of the hydrophobic nature of PNIPAM particles with increasing temperature \cite{Romeo2010}.
\paragraph{}
One of the highlights of this work is the identification of a gel-liquid-gel transition when the temperature of a dense PNIPAM suspension constituted by soft particles was raised across the LCST. In results that verify our rheology data, we simultaneously observed that suspensions of soft PNIPAM particles form gel-like networks at temperatures below and above the LCST, while suspensions of stiff PNIPAM particles remain in a glassy state at all the experimental temperatures studied here. We show here, for the first time to the best of our knowledge, that phase transformations in thermoreversible microgel suspensions depend sensitively on the stiffness of the constituent particles. By tuning just the PNIPAM particle stiffness and suspension temperature, we have demonstrated here that a rich variety of microgel suspension phases with unique properties can be achieved. In future studies, it would be interesting to use osmotic pressure measurements to find a quantitative relation between the real and effective volume fractions of PNIPAM suspensions constituted by particles of different stiffnesses, such as implemented by Scotti $et$ $al.$ \cite{A_scotti_PRE_2021}. Additionally, the dependence of bulk rheology and phase behavior on the dynamics of ions in microgel suspensions and the swelling properties of PNIPAM particles in different ionic solvents are interesting topics that need to be investigated. It would also be interesting to study the role of a non-polar crosslinker on the phase transformations and rheology of dense suspensions of microgel particles.
\paragraph{}
Since the mechanical properties and microstructures of dense PNIPAM suspensions can be controlled simply by tuning the suspension temperature and stiffness of the constituent particles, their aqueous suspensions are potentially good candidates for synthesizing multifunctional materials. The tunability of suspension rheology by controlling particle stiffness and medium temperature can be exploited in the design of soft robotic applications, for example, in constructing microgrippers, folding flat composite gel sheets \cite{W_J_Zheng_2015} and in soft biomimetic machines. The bending or twisting of a flat gel sheet, comprising layers of microgel particle aggregates of different stiffnesses can, in principle, be achieved by the application of a temperature gradient. PNIPAM microgel suspensions are extremely resilient to tensile and compressive stresses \cite{Hu_Materials-design_2019} and are also very flexible. Since these materials can be used in various engineering and biomimetic applications, further research on their mechanical properties is of utmost importance.

\section*{Author Contributions}
\textbf{Chandeshwar Misra:} Methodology, Validation, Formal analysis, Investigation, Writing - Original Draft preparation, Visualization. \textbf{Sanjay Kumar Behera:} Methodolgy. \textbf{Sonali Vasant Kawale:} Methodolgy. \textbf{Ranjini Bandyopadhyay:}
Conceptualization, Methodology, Validation, Resources, Data curation, Writing - Review \& Editing, Supervision, Project administration, Funding acquisition.

\section*{Conflicts of interest}
The authors have no conflicts to disclose.

\section*{Acknowledgements}
The authors thank K. M. Yatheendran for his help with cryo-SEM imaging and K. N. Vasudha for her help with FTIR and DSC measurements. We thank Department of Science and
Technology, Science and Engineering Research Board (grant
number EMR/2016/006757) for partial financial support.





\bibliographystyle{ieeetr}
\bibliography{Ref_stiff_SM.bbl}
\includepdf[pages=-]{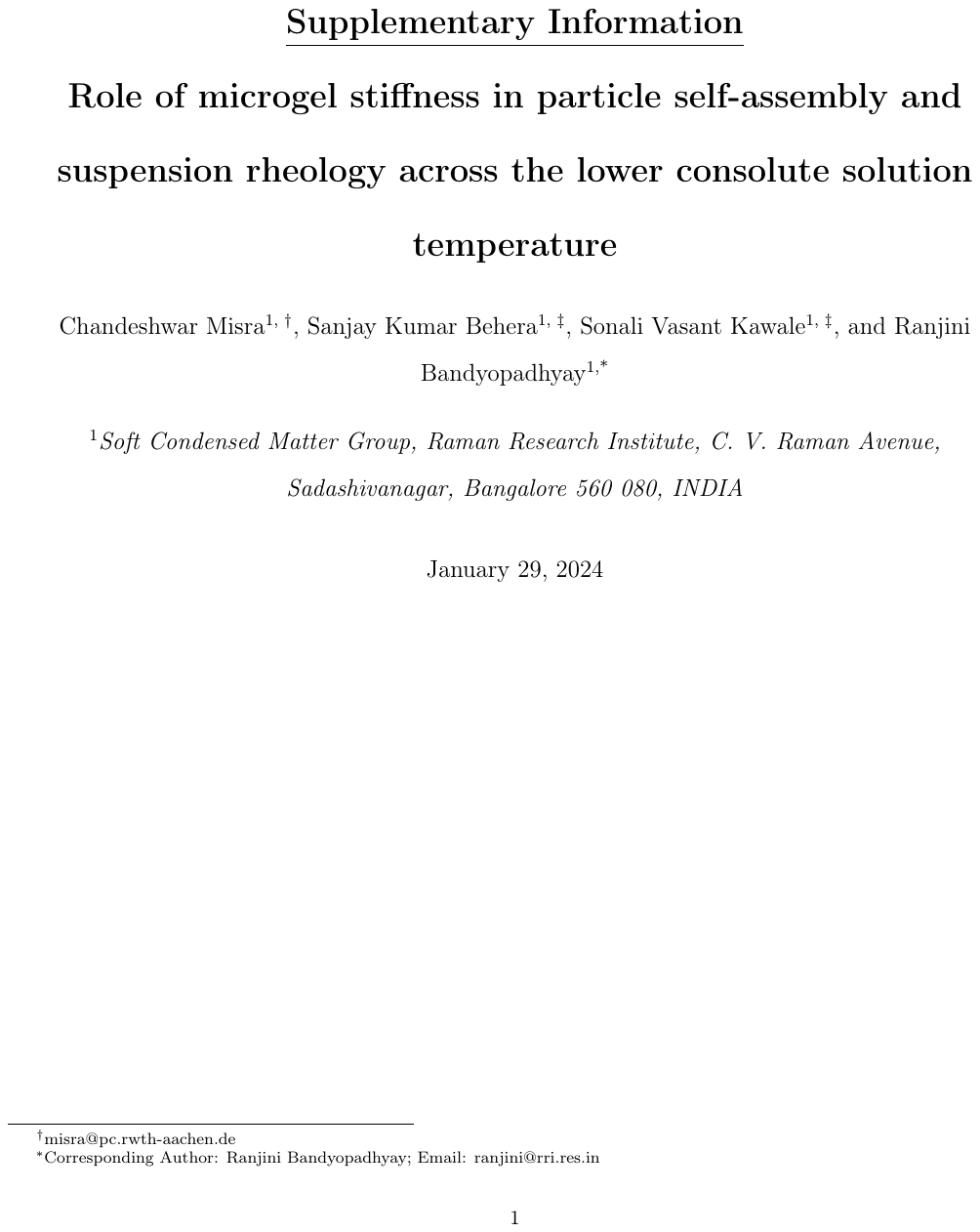}
\end{document}